\documentclass[twocolumn,aps]{revtex4}
\usepackage{graphicx}

\newcommand{\one}{\leavevmode\hbox{\small1\normalsize\kern-.33em1}}

\begin{document}

\title{Estimating Pre- and Post-Selected Ensembles}

\author{Serge Massar}

\affiliation{Laboratoire d'Information Quantique, {C.P.} 225, Université Libre
de Bruxelles (U.L.B.), Boulevard du Triomphe, B-1050 Bruxelles, Belgium}

\author{Sandu Popescu}

\affiliation{H. H. Wills Physics Laboratory, University of Bristol, Tyndall Avenue,
Bristol BS8 1TL, U.K., and Hewlett-Packard Laboratories, Stoke Gifford,
Bristol BS12 6QZ, U.K.}

\date{\today}
\begin{abstract}
In analogy with the usual quantum state estimation problem, we introduce
the problem of state estimation for a pre- and post-selected ensemble.
The problem has fundamental physical significance since, as argued
by Y. Aharonov and collaborators, pre- and post-selected ensembles
are the most basic quantum ensembles. Two new features are shown to
appear: 1) information is flowing to the measuring device both from
the past and from the future; 2) because of the post-selection, certain
measurement outcomes can be forced never to occur. Due to these features,
state estimation in such ensembles is dramatically different from
the case of ordinary, pre-selected only ensembles. We develop a general
theoretical framework for studying this problem, and illustrate it
through several examples. We also prove general theorems showing the
intimate relation between information flowing from the future is related
to the complex conjugate information flowing from the past. Finally
we illustrate our approach on examples involving covariant measurements
on spin 1/2 particles. We emphasize that \textit{all} state estimation
problems can be extended to the pre- and post-selected situation.
The present work thus lays the foundations of a much more general
theory of quantum state estimation.
\end{abstract}
\maketitle

\section{Introduction}

Quantum mechanics is usually formulated in terms of initial conditions.
The state $|\psi_{i}\rangle$ is given at time $t_{i}$ and then evolves
according to the Schr{ö}dinger equation. However it was realized
in \cite{ABL} that one could also use a time symmetric formulation
in which one imposes both the initial condition $|\psi_{i}\rangle$
at the initial time $t_{i}$ and the final condition $\langle\psi_{f}|$
at final time $t_{f}$. For an exposition we refer to the review \cite{AV}
and to \cite{APTV} where the concept of pre- and post-selection has
been extended to multiple time states. Pre- and post-selection gives
rise to a number of paradoxes and surprising effects that do not occur
in the standard formulation of quantum theory. Studying them is a
worthy endeavor: pre- and post-selected ensembles are \textit{the}
most detailed quantum ensemble one can prepare, hence arguably they
are \textit{the} fundamental quantum ensembles.

Independently of the above line of work, the past decades have seen
the development of quantum information theory, and in particular an
in depth study of quantum state estimation, see e.g. \cite{Helstrom,Holevo,MP,GP,M,Betal1,Betal2,Betal3}.
The general problem of state estimation is, given an unknown quantum
state $\psi$, to devise the best procedure to estimate the state.

In the present paper we try to bring together these two lines of inquiry.
We consider the problem of estimating an unknown ensemble, when both
the pre- and the post-selected states are unknown. This differs from
the usual state estimation problem because information is flowing
to the observer both from the past and from the future. In the first
part of the paper we will show how to formulate this problem. Two
new features are shown to appear: 1) information is flowing to the
measuring device both from the past and from the future; 2) because
of the post-selection, certain measurement outcomes can be forced
never to occur. Due to these features, state estimation in such ensembles
is very different from the case of ordinary, pre-selected only ensembles.
For instance, in the usual state estimation problem in which information
arrives only from the past, measurements are described by Positive
Operator Valued Measures (POVM), whereas when information arrives
both from the past and from the future, measurements are described
by Kraus operators. In a second part of the paper, we prove general
theorems establishing that information flowing from the future and
the complex conjugate information flowing from the past are closely
related, and in some cases equivalent. In the final part of the paper,
we illustrate this formalism on examples involving covariant measurements
on spin 1/2 particles.

Considerable work has already been devoted to studying measurements
on pre- and post-selected ensembles. These works have mainly focused
on the counterintuitive results which can be exhibited by ``weak
measurements'' carried out at an intermediate time, between
fixed pre- and post-selected states\cite{ref1}. This approach has
applications for understanding quantum paradoxes, see for instance
the experiments \cite{ref4} on Hardy's paradox \cite{ref5,ref6},
and the recent experiments measuring wave functions and trajectories
of quantum particles \cite{refLSPSB,refKBRSMSS}); for describing
superluminal light propagation\cite{ref7,ref8}; for computing polarization
mode dispersion effects in optical networks \cite{ref9}; as well
as experiments in cavity QED \cite{ref10}. Other experimental investigations
of weak measurements are reported in \cite{ref11}. In addition it
was shown, following the initial suggestion of \cite{ref13}, that
weak measurements can have applications for high precision measurements.
These include the first observation of the spin Hall effect \cite{ref14}
and the observation of small transverse deflections of a light beam
\cite{ref15}; see also the proposals for measurements of charge \cite{refZRG}
and of imaginary phase shifts \cite{refBS}. Note that in all these
works the pre- and post-selected states are kept fixed, and it is
the effects of the measurement which are investigated.

Closely related to the present work is \cite{A} where it was shown
that in the presence of a fixed post-selected state, some (pre-selected)
states can be estimated to extremely high precision, with as consequence
that the computational power of pre- and post-selected ensembles is
equivalent to the complexity class PP. This shows that the presence
of a post-selected state can dramatically change the state estimation
problem because certain measurement outcomes can be forced never to
occur. Another well known example which can be interpreted in the
same way (see discussion below) is the Unambiguous State Estimation
(USE) problem\cite{USD1,USD2,USD3}.

Here we both formulate in full generality the problem of state estimation
in the presence of a post-selected state, and introduce the new problem
of estimating an unknown pre- and post-selected ensemble. At this
stage we do not know if this approach will have applications (e.g.
for high precision measurements), rather in this first work we are
interested in the conceptual issue of formulating this problem and
understanding its relation to the usual state estimation problem.

\section{Setting up the problem}

\label{settingup} 

\subsection{The standard state estimation problem}

\begin{figure}
{\includegraphics[height=80mm]{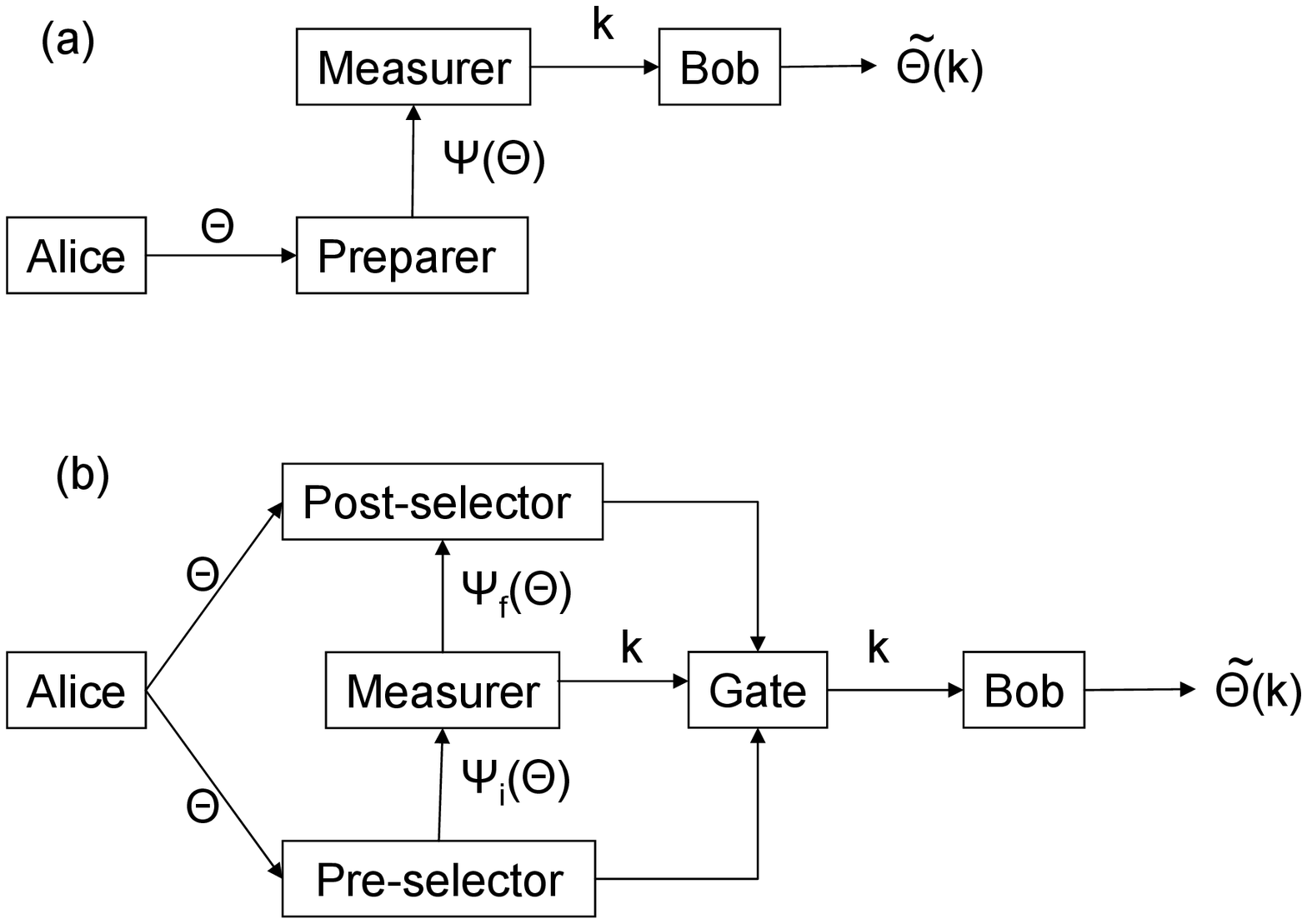}} \caption{State Estimation. Panel (a) illustrates the usual formulation of state
estimation. The parameter $\theta$ is chosen by Alice. It determines
the quantum state $|\psi(\theta)\rangle$ which is prepared by the
Preparer. The state is then sent to the Measurer who carries out a
quantum measurement, obtaining outcome $k$. The outcome is then sent
to Bob who, on the basis of the value $k$, tries to guess what was
the value of the parameter $\theta$. His guess is denoted $\tilde{\theta}(k)$.
Panel (b) illustrates the estimation of pre- and post-selected ensemble.
The parameter $\theta$ is chosen by Alice. It determines the pre-
and post-selected states $|\psi_{i}(\theta)\rangle$ and $\langle\psi_{f}(\theta)|$.
These are prepared by the Pre- and Post-selectors respectively. The
Measurer carries out a quantum measurement at an intermediate time,
obtaining outcome $k$. The role of the Gate is to check if the pre-
and post-selections succeeded. Only if the pre- and post-selections
succeeded is the outcome $k$ of the measurement sent to Bob. Upon
receiving $k$, Bob tries to guess what was the value of the parameter
$\theta$. His guess is denoted $\tilde{\theta}(k)$. }

\label{figure:one}
\end{figure}

It will be useful to view the standard state estimation problem as
a game played between Alice and Bob: Alice chooses a parameter $\theta$
and Bob must try to guess the value of $\theta$, given access to
a quantum state $|\psi(\theta)\rangle$. We call $\tilde{\theta}$
Bob's guess, which should be as close as possible (according to some
merit function $F$) to the true value $\theta$. We now define this
game with precision. To this end we introduce several additional actors
that follow the instructions of either Alice or Bob. The whole state
estimation problem consists of the following steps (described graphically
in panel (a) of figure \ref{figure:one}):

1) Alice chooses a (multidimensional) parameter $\theta$ taken from
some set $\Theta$ according to a probability distribution $p(\theta)$.
The set $\Theta$ and probability distribution $p(\theta)$ are known
to Bob.

2) The first actor is the Preparer. He receives from Alice the value
of $\theta$ and prepares a quantum state $|\psi(\theta)\rangle$.
The dependency of the quantum state on the parameter $\theta$ (i.e.,
the function $|\psi(\theta)\rangle$) is known to Bob.

3) The second actor is the Measurer who carries out a measurement
on the state provided by the Preparer. The POVM is chosen by Bob.
Denote the outcome of the measurement by $k$. The Measurer sends
the value of $k$ to Bob.

4) Finally Bob outputs a guess $\tilde{\theta}(k)$ which depends
on the value of $k$. The quality of the guess is measured by some
merit function $F(\theta,\tilde{\theta}(k))$.

The experiment is then repeated many times. Each time Alice chooses
a new value for $\theta$ according to the probability distribution
$p(\theta)$. The quality of the state estimation procedure is measured
by the average of the merit function $F$.

The above scenario may seem overly complicated. However the separation
of the roles of the different actors will become important in the
pre- and post-selected case.

Note that here and throughout this manuscript we neglect the free
(unitary) evolution between preparation and measurement. Any such
free evolution is supposed to be known to the parties, and can therefore
be taken into account. 

\subsection{Estimating a Pre- and Post-Selected ensemble}

We now setup, in parallel with the standard state estimation problem,
the problem of estimating a pre- and post-selected ensemble.

First of all note that, although the aim of pre- and post-selection
is to have a formulation in which past and future play symmetric roles,
it is often useful to rephrase the problem in the language of usual
quantum mechanics, in which the past and future play nonequivalent
roles. Then by imposing that the post-selection succeeds, one recovers
a time symmetric formulation. In the following paragraphs we take
this more traditional point of view.

Once more, we view the estimation of a pre- and post-selected ensemble
as a game played between Alice and Bob: Alice chooses a parameter
$\theta$ and Bob must try to guess the value of $\theta$, given
access to a pre- and post-selected ensemble $\langle\psi_{f}(\theta)||\psi_{i}(\theta)\rangle$.
We call $\tilde{\theta}$ Bob's guess, which should be as close as
possible (according to some merit function $F$) to the true value
$\theta$. We now define this game with precision. To this end we
introduce several additional actors. The estimation problem consists
of the following steps (described graphically in panel (b) of figure
\ref{figure:one}):

1) Alice chooses a (multidimensional) parameter $\theta$ taken from
some set $\Theta$ according to a probability distribution $p(\theta)$.
The set $\Theta$ and probability distribution $p(\theta)$ are known
to Bob. She sends the value of $\theta$ to the Pre-selector and to
the Post-selector (see steps 2 and 4).

2) The first actor is the Pre-selector. He prepares a quantum state
$|\psi_{i}(\theta)\rangle$. The dependency of the quantum state on
the parameter $\theta$ (i.e., the function $|\psi_{i}(\theta)\rangle$)
is known to Bob.

3) The second actor is the Measurer who carries out a measurement
on the state provided by the Pre-selector. The actions of the Measurer
are chosen by Bob. The result of the measurement consists of two pieces.
First of all the classical data produced by the measurement. Call
this classical information $k$. Second the quantum state of the system,
modified by the action of the measurement. After the measurement is
finished, the Measurer sends the classical information $k$ to a logical
Gate (see step 5) and sends the quantum state of the system to the
Post-selector (see step 4).

4) The third actor is the Post-selector. He checks whether or not
the state sent to him by the Measurer is $\langle\psi_{f}(\theta)|$.
He does this by measuring an observable that has $\langle\psi_{f}(\theta)|$
as one of its non degenerate eigenstates. He sends the result of his
measurement to the Gate (see step 5). The dependency of the quantum
state on the parameter $\theta$ (i.e., the function $\langle\psi_{f}(\theta)|$)
is known to Bob.

5) The Gate receives the value $k$ from the Measurer, and the information
on whether the post-selection succeeded from the Post-selector. If
the post-selection succeeded, then the Gate sends the result $k$
of the measurement to Bob. If the post-selection failed, then the
Gate instructs the Pre-Selector, Post-Selector, and Measurer that
they must start over at step 2, the value of $\theta$ being kept
fixed.

6) Finally Bob outputs a guess $\tilde{\theta}(k)$ which depends
on the value of $k$. The quality of the guess is measured by some
merit function $F(\theta,\tilde{\theta}(k))$.

In the present work we are interested in the information contained
in the pre- and post-selected ensemble itself, i.e. in the \textit{conditional}
information, given that we succeeded to prepare the ensemble. We want
to exclude that information about the probability to actually prepare
the ensemble can be used to estimate the ensemble. The role of the
Gate in the above procedure is to make this condition explicit. Indeed,
because of the Gate, Bob only receives the result of the measurement
if the pre- and post-selection succeeded and does not have any information
on how many times step 3 must be repeated before the post-selection
succeeds.

Note that one can consider the case where the Post-selector post-selects
a fixed state $\langle0|$ which does not depend on $\theta$, or
a combination $\langle\psi_{f}(\theta)|\langle0|$ of a state which
depends on $\theta$ and a state that does not. We refer to these
situations as the cases where there is a {}``fixed post-selected
state''.

Note that although the above setup is described within the usual framework
of quantum theory, with evolution going from the past to the future,
the final expressions for the quality of the ensemble estimation by
Bob will be time-symmetric. The pre- and post-selected states will
play the same role. This will become apparent below.

\section{State estimation in the presence of a fixed post-selected state}

\label{sec:fixed}

Before studying the general case, it is useful to consider a simple
situation, namely the case in which the post-selected state is fixed
(i.e. independent of $\theta$). Indeed this case is closest to the
usual state estimation problem, and several interesting results have
already been obtained in the literature which can help develop an
intuition. For definiteness we denote the fixed post-selected state
$\langle0|$. When the post-selected state is fixed no information
flows to the Measurer from the future - there simply is no information
in the post-selected state since there is no uncertainty about it.
So naively, one would expect that in this case the estimation problem
is identical to the standard pre-selected only case. However, as we
now show, the existence of a post-selected state completely changes
the state estimation problem.

One way to interpret this situation is that the Measurer can reject
certain measurement outcomes for free. Namely, if the measurement
provides a useful outcome, the Measurer prepares the state $|0\rangle$,
and the post-selection will succeed. On the contrary, if the measurement
outcome is not useful, the Measurer prepares the state $|1\rangle$,
the post-selection will fail, and he will be allowed to begin the
measurement anew on a fresh copy of the state. 

In this context, a dramatic example is provided by the problem of
Unambiguous State Estimation\cite{USD1,USD2,USD3}. Suppose the Pre-selector
prepares one of two non orthogonal states $|\psi_{1}\rangle$ and
$|\psi_{2}\rangle$, while the Post-selector selects the fixed state
$\langle0|$. The task of Bob is to say either ``the state is $\psi_{1}$'',
or ``the state is $\psi_{2}$'', or ``I do not
know''. The constraint is that if one says that the state
is $\psi_{1}$ ($\psi_{2}$), then one cannot make a mistake.

As is well known, in the standard unambiguous state discrimination
problem (i.e. without post-selection) such discrimination is possible
for all pairs of states $\psi_{1}$ and $\psi_{2}$, but the probability
of success goes to zero as the states $|\psi_{1}\rangle$ and $|\psi_{2}\rangle$
get closer and closer $|\langle\psi_{1}|\psi_{2}\rangle|\to1$. However,
in presence of post-selection Bob can always succeed. The procedure
is for the Measurer to perform the standard (pre-selected only) unambiguous
state discrimination and then prepare the system in the state $|0\rangle$
whenever the measurement indicates $\psi_{1}$ or $\psi_{2}$, but
prepare the system in state $|1\rangle$ whenever the outcome is ``I
do not know''. The ``I do not know''
cases will thus never pass post-selection and will never be counted.

A second spectacular example is taken from \cite{A}. Suppose that
the Preparer prepares $n$ identical particles all prepared in the
same state $|\uparrow_{\theta}\rangle=\cos\theta/2|\uparrow\rangle+\sin\theta/2|\downarrow\rangle$.
We are promised that either $\theta\in S_{+}=[\epsilon,\pi-\epsilon]$
or $\theta\in S_{-}=[-\pi+\epsilon,-\epsilon]$ with $0<\epsilon<\pi/2$.
The task is to distinguish whether $\theta$ belongs to set $S_{+}$
or to set $S_{-}$. We are allowed a small error probability (say
$P(error)<1/3$). If the Measurer is promised that there is a fixed
post-selected state $\langle0|$, then this task can be solved with
$n=O((\log1/\epsilon)\log(\log1/\epsilon))$ particles. On the other
hand in the usual formulation of state estimation with no post-selection
one needs $n=O(\epsilon^{-2})$ particles. This is a huge difference
and has dramatic consequences: essentially the same state estimation
problem is used in \cite{A} to show that a quantum computer with
access to a post-selected state $\langle0|$ can solve PP complete
problems. 

More generally one can take any state estimation problem in the standard
formulation, and inquire how the quality of the state estimation changes
if there is a fixed post-selected state $\langle0|$. Does one always
get dramatic improvements as in the above two examples? Below we will
analyze the cases of covariant measurements on $n$ spin 1/2 particles
in the state $|\uparrow_{\Omega}^{\otimes n}\rangle$, and of covariant
measurements when the $n$ spins are in the state $|\uparrow_{\Omega}^{\otimes n/2}\downarrow_{\Omega}^{\otimes n/2}\rangle$
(for $n$ even). We will see that in these cases the presence of a
fixed post-selected state $\langle0|$ can sometimes give a small
increase in Fidelity, but nothing as spectacular as in the above examples.

Before presenting these results, we first give a general framework
for describing state estimation in the pre- and post-selected context.

\section{General Formalism}

We give here general expressions for state estimation in the presence
of both pre- and post-selection. In section \ref{sec:Interaction-between-a}
we argue why these are the natural generalizations of the standard
formalism.

\subsection{Standard state estimation\label{sub:Standard-state-estimation}}

For definiteness we first recall standard state estimation theory.
In this case the most general measurement is a Positive Operator Valued
Measure (POVM) described by operators $M_{k}$ which are positive
and sum to the identity: 
\begin{equation}
M_{k}\geq0\quad,\quad\sum_{k}M_{k}=\one\ .\label{povm}
\end{equation}
 The probability of finding outcome $k$ if the state was $|\psi\rangle$
is 
\begin{equation}
P(k|\psi)=\langle\psi|M_{k}|\psi\rangle\ .
\end{equation}
 The average value of the merit function can then be expressed as
\begin{equation}
F=\int\! d\theta p(\theta)\sum_{k}\langle\psi(\theta)|M_{k}|\psi(\theta)\rangle F(\theta,\tilde{\theta}(k))\ .\label{F1-S}
\end{equation}
We note the well known fact that POVMs with rank one operators are
the most informative (for a proof see the argument at the end of section
\ref{sub:State-est-with-fixed-postselectedstate}).

\subsection{State estimation with a fixed post-selected state.\label{sub:State-est-with-fixed-postselectedstate}}

\label{IVB}

When there is a fixed post-selected state (the situation discussed
in section \ref{sec:fixed}) the preceding formalism must be generalized.
The probability of finding outcome $k$ if the state was $|\psi\rangle$
is 
\begin{equation}
P_{M}(k|\psi)=\frac{\langle\psi|M_{k}|\psi\rangle}{\sum_{k'}\langle\psi|M_{k'}|\psi\rangle}\label{eq:PMfixedpost}
\end{equation}
 where the operators $M_{k}$ are positive but no longer normalized
(we say they are {}``subnormalized''): 
\begin{equation}
M_{k}\geq0\quad,\quad\sum_{k}M_{k}\leq\one\ .\label{povm-post}
\end{equation}
 The average value of the merit function can then be expressed as
\begin{equation}
F=\int\! d\theta p(\theta)\sum_{k}\frac{\langle\psi(\theta)|M_{k}|\psi(\theta)\rangle}{\sum_{k'}\langle\psi(\theta)|M_{k'}|\psi(\theta)\rangle}F(\theta,\tilde{\theta}(k))\ .\label{F1-S2}
\end{equation}

We now show that \textit{POVMs with rank one operators are the most
informative}, whether or not there is a fixed post-selected state.
Consider an arbitrary POVM with elements $M_{k}$ and associated estimator
$\tilde{\theta}(k)$. Since the $M_{k}\geq0$ are positive operators,
we can write them as $M_{k}=\sum_{j}|m_{kj}\rangle\langle m_{kj}|$,
with $|m_{kj}\rangle$ unnormalised states. Consider the refined POVM
with elements $M_{kj}=|m_{kj}\rangle\langle m_{kj}|$. If to the refined
POVM element $M_{kj}$ we associate the same estimator $\tilde{\theta}(k)$
as for the original POVM, then the value of the merit function does
not change (to see this note that the denominator in eq. (\ref{F1-S2})
does not change when one replaces the original POVM by the refined
POVM with elements $M_{kj}$). Hence the value of the merit functions
for POVMs with rank one elements are always at least as large as the
merit functions for unrefined POVMs.

\subsection{Estimation of pre- and post-selected ensembles}

\label{IVC}

In the case of estimation of pre- and post-selected ensembles, the
measurement operators are no longer POVM elements, but Kraus operators.
Kraus operators describe the most general evolution of an open quantum
system: 
\begin{equation}
\rho\to\sum_{k}A_{k}\rho A_{k}^{\dagger}
\end{equation}
 and are normalized according to 
 \begin{equation}
 \sum_{k}A_{k}^{\dagger}A_{k}=\one\ .
 \label{normKraus}
 \end{equation}

Kraus operators are the appropriate operators to describe interaction
with a pre- and post-selected ensemble because the Kraus operators
consist of a ket-bra which points both towards the past and towards
the future: 
\begin{equation}
A_{k}=\sum_{l}|\phi_{k}^{l}\rangle\langle\varphi_{k}^{l}|\ .
\end{equation}
 In addition, if there is a fixed post-selected state, one must also
modify the normalization condition, and replace the equality in eq.
(\ref{normKraus}) by an inequality (we say the Kraus operators are
``subnormalized'').

We thus have for the probability of obtaining outcome $k$ conditional
on the pre- and post-selected states $\langle\psi_{f}||\psi_{i}\rangle$:
\begin{equation}
P_{A}(k|\psi_{f},\psi_{i})=\frac{|\langle\psi_{f}|A_{k}|\psi_{i}\rangle|^{2}}{\sum_{k'}|\langle\psi_{f}|A_{k'}|\psi_{i}\rangle|^{2}}\ .\label{Pkprepost}
\end{equation}
 with the normalization 
\begin{eqnarray}
\sum_{k}A_{k}^{\dagger}A_{k} & = & \one\quad\mbox{no additional post-selection}\\
\mbox{or}\nonumber \\
\sum_{k}A_{k}^{\dagger}A_{k} & \leq & \one\quad\mbox{fixed post-selected state}
\end{eqnarray}
 Note that in this expression $\langle\psi_{f}|$ need not belong
to the same Hilbert space as $|\psi_{i}\rangle$, as Kraus operators
allow to describe the evolution of a system belonging to one Hilbert
space into a system belonging to another Hilbert space. The average
value of the merit function can then be expressed as 
\begin{equation}
F=\int\! d\theta p(\theta)\sum_{k}\frac{|\langle\psi_{f}(\theta)|A_{k}|\psi_{i}(\theta)\rangle|^{2}}{\sum_{k'}|\langle\psi_{f}(\theta)|A_{k'}|\psi_{i}(\theta)\rangle|^{2}}F(\theta,\tilde{\theta}(k))\ .\label{F1}
\end{equation}

\section{Interaction between a system and a measuring device\label{sec:Interaction-between-a}}

We now go back to the setups presented in section \ref{settingup},
and argue why the expressions given in sections \ref{IVB} and \ref{IVC}
are a natural generalization of the Born rule to the case of pre-
and post-selected ensembles. Note that we cannot provide a proof that
they constitute the only possible generalization, but only plausibility
arguments.

To derive the above expressions for the probability of obtaining the
outcome $k$, we go back to the general setup described in section
\ref{settingup} and Fig. \ref{figure:one}, and describe it in the
standard quantum formalism.

To simplify the problem, let us first note that if there is a fixed
post-selected state, then we can without loss of generality take it
to be a single qubit. Indeed in this case the most general procedure
is for the Post-Selector to post select that the state belongs to
a subspace. Denote by $\Pi$ the projector onto this subspace and
by $|\Psi\rangle$ the state just before projection onto $\Pi$. We
now describe an equivalent post-selection in which only a single qubit
is used. First we add to the system space an ancillary qubit initially
in the state $|0\rangle$. The state is thus $|\Psi\rangle\otimes|0\rangle$.
Next we carry out the unitary 
\[
U=\Pi\otimes|0\rangle\langle0|+(\one-\Pi)\otimes\left(|1\rangle\langle0|+|0\rangle\langle1|\right)\ ,
\]
 i.e. a controlled NOT, where the projection onto $\Pi$ acts as control.
Finally we post-select that the qubit is in state $|0\rangle$. The
probability of success of this post-selection is exactly the same
as the original one, hence the two methods are equivalent.

We now go back to the general setup described in section \ref{settingup}
and Fig. \ref{figure:one}. Let $|\psi_{i}\rangle_{S}\in H_{S}$ be
the initial state of the system. We adjoin to $H_{S}$ two additional
Hilbert spaces. First there is the Hilbert space $H_{R}$ of the measurement
register. The initial state of the measurement register is $|0\rangle_{R}$.
If the final state is $|k\rangle_{R}$, then the outcome of the measurement
will be $k$. Second there is the Hilbert space of single qubit $H_{P}$
which is used in case there is a fixed post-selection. The initial
state of this qubit is $|0\rangle_{P}$. The fixed post-selection
will succeed if the final state of this qubit is still $|0\rangle_{P}$.
The initial state is thus 
\begin{equation}
|\psi_{i}\rangle_{S}\otimes|0\rangle_{R}\otimes|0\rangle_{P}
\end{equation}
 where the subscripts denote to which Hilbert space each state belongs.
The action of the Measurer can be described by a unitary evolution
$U$ that entangles the Hilbert spaces $H_{S}$, $H_{R}$, $H_{P}$.
This yields the state 
\begin{eqnarray}
 & U|\psi_{i}\rangle_{S}\otimes|0\rangle_{R}\otimes|0\rangle_{P}\\
 & =\sum_{k}\sum_{x=0,1}\bigl[\one_{S}\otimes|k\rangle_{R}\langle k|\otimes|x\rangle_{P}\langle x|\bigr]\nonumber \\
 & \quad\quad\quad\times U|\psi_{i}\rangle_{S}\otimes|0\rangle_{R}\otimes|0\rangle_{P}\\
 & =\sum_{k}\sum_{x=0,1}\bigl(A_{kx}|\psi_{i}\rangle_{S}\bigr)\otimes|k\rangle_{R}\otimes|x\rangle_{P}
\end{eqnarray}
 where unitarity of $U$ imposes that 
\begin{equation}
\sum_{kx}A_{kx}^{\dagger}A_{kx}=\one_{S}
\end{equation}

Consider first the case where the only post-selected state is the
fixed state $_{P}\langle0|$. The probability to find the register
in state $k$ and for the post-selection to succeed is 
\begin{equation}
P(k,x=0|\psi_{i})=\langle\psi_{i}|A_{k0}^{\dagger}A_{k0}|\psi_{i}\rangle
\end{equation}
 Because of the presence of the Gate that checks that the post-selection
succeeded, the relevant quantity is the probability to find the register
in state $k$ {\em conditional} on the post-selection having succeeded.
This is 
\begin{eqnarray}
P(k|\psi_{i},x=0) & = & \frac{\langle\psi_{i}|A_{k0}^{\dagger}A_{k0}|\psi_{i}\rangle}{\sum_{k'}\langle\psi_{i}|A_{k'0}^{\dagger}A_{k'0}|\psi_{i}\rangle}\nonumber \\
 & = & \frac{\langle\psi_{i}|M_{k}|\psi_{i}\rangle}{\sum_{k'}\langle\psi_{i}|M_{k'}|\psi_{i}\rangle}
\end{eqnarray}
 where $M_{k}=A_{k0}^{\dagger}A_{k0}$ are POVM elements: they are
hermitian, positive $M_{k}\geq0$, and bounded by the identity: $\sum_{k}M_{k}\leq\one$.
We thereby obtain the formalism of section \ref{IVB}.

Consider now the case where one post-selects both that the final state
is $_{S}\langle\psi_{f}|$ and that there is the fixed post-selected
state $_{P}\langle0|$. The amplitude of finding state $_{S}\langle\psi_{f}\otimes_{R}\langle k|\otimes_{P}\langle0|$
is $\langle\psi_{f}|A_{k0}|\psi_{i}\rangle$. The probability of this
event is 
\begin{equation}
P(k,x=0,\psi_{f}|\psi_{i})=|\langle\psi_{f}|A_{k0}|\psi_{i}\rangle|^{2}
\end{equation}
 Because of the presence of the Gate that checks that the post-selection
succeeded, the relevant quantity is the probability to find the register
in state $k$ {\em conditional} on the post-selections having succeeded.
This is 
\begin{equation}
P(k|\psi_{i},\psi_{f},x=0)=\frac{|\langle\psi_{f}|A_{k0}|\psi_{i}\rangle|^{2}}{\sum_{k'}|\langle\psi_{f}|A_{k'0}|\psi_{i}\rangle|^{2}}
\end{equation}
 where the operators $A_{k0}$ are arbitrary, except for the condition
$\sum_{k}A_{k0}^{\dagger}A_{k0}\leq\one_{S}$.

Note that if there is no fixed post-selection onto $_{P}\langle0|$,
then the above calculation carries through with the Hilbert space
$H_{P}$ (and hence the index $x$) omitted. One then obtains the
standard normalization for the Kraus operators $\sum_{k}A_{k}^{\dagger}A_{k}=\one_{S}$%
\footnote{In some cases, the post-selection of a state $\langle\psi_{f}|$ by
itself implies the existence of a fixed post-selection. For instance
suppose that $\langle\psi_{f}|=\alpha\langle0|+\beta\langle0|$ belongs
to a two dimensional subspace of a three dimension space with basis
$\langle0|,\langle1|,\langle2|$. Then whenever the Measurer does
not want an outcome to occur, he prepares the state $|2\rangle$,
and the post-selection never occurs. On the other hand it may be that
the Hilbert space to which belongs $\langle\psi_{f}|$ is intrinsically
two dimensional. (For instance polarization of a photon). In this
case there is a difference between the presence or not of a fixed
post-selection. For this reason we keep the two notions distinct in
the present paper.%
}.

Note that if the post-selected state $\langle\psi_{f}|$ belongs to
a different Hilbert space then the pre-selected state $|\psi_{i}\rangle$,
then the above calculation carries through unchanged, except that
one must enlarge the Hilbert space to contain both the spaces of the
initial state and that of the final state.

The above analysis thus leads to the formalism of section \ref{IVC}.

\section{Information flow from the past and from the future}

\subsection{Two Theorems.}

How well can we estimate the parameter $\theta$ in the above situations?
Obviously the estimation can be done better in the pre- and post-selected
ensemble than if one is given the pre-selected state $|\psi_{i}(\theta)\rangle$
only, since the post-selected state provides additional information.
But how much more information? We now show that the relevant comparison
is with the pre-selected tensor product state $|\overline{\psi_{f}(\theta)}\rangle\otimes|\psi_{i}(\theta)\rangle$,
where $|\overline{\psi}\rangle$ is the state obtained by complex
conjugating the coefficients of $|\psi\rangle$ in a basis: $|\psi\rangle=\sum_{k}c_{k}|k\rangle\to|\overline{\psi}\rangle=\sum_{k}\overline{c_{k}}|k\rangle$.

Some intuition for this mapping can be obtained by recalling that
in a pre- and post-selected ensemble, the pre-selected state arrives
from the past, whereas the post-selected state arrives from the future.
It is thus natural that it behaves like the time reverse of a pre-selected
state. And time reversal is realized mathematically by complex conjugation.
Another motivation follows from the remark made in \cite{APTV} that
it is possible to realize a pre- and post-selected ensemble $\langle\psi_{f}(\theta)|$
$|\psi_{i}(\theta)\rangle$ by: A) pre-selecting the tensor product
state $|\overline{\psi_{f}(\theta)}\rangle\otimes|\psi_{i}(\theta)\rangle$,
B) post-selecting the maximally entangled state $\sum_{k}\langle k|\otimes\langle k|$
(where $|k\rangle$ is the basis in which the complex conjugation
is defined), and C) at intermediate times acting only on the second
system.

Thus both lines of reasoning suggest that to a post-selected state
$\langle\psi_{f}|$ we should associate the pre-selected complex conjugate
state $|\overline{\psi}_{f}\rangle$. The following results put this
intuition on a firm basis. To state them we use the following notation. 

Denote by $|\phi\rangle\in H^{d}$ and $|\psi\rangle\in H^{d'}$ states
belonging to Hilbert spaces of dimension $d$ and $d'$ respectively.
Denote by $|\overline{\phi}\rangle$ the state obtained from $|\phi\rangle$
by complex conjugation in a (fixed but arbitrary) basis. Consider
a subnormalised POVM acting on the tensor product space $H^{d}\otimes H^{d'}$
with rank one elements: $M_{k}=|m_{k}\rangle\langle m_{k}|$, $\sum_{k}M_{k}\leq\one$.
The probability of outcome $k$ when the state is the tensor product
$|\overline{\phi}\rangle\otimes|\psi\rangle$ is given by eq. (\ref{eq:PMfixedpost}):
\begin{equation}
P_{M}(k|\overline{\phi}\psi)=\frac{|\langle m_{k}|\overline{\phi}\rangle\otimes|\psi\rangle|^{2}}{\sum_{k'}|\langle m_{k}|\overline{\phi}\rangle\otimes|\psi\rangle|^{2}}\ .\label{eq:TheoremPM-1}
\end{equation}
Consider a subnormalised Completely Positive Map described by Kraus
operators $A_{k}:H^{d'}\to H^{d}$, $\sum_{k}A_{k}^{\dagger}A_{k}\leq\one$.
The probability of finding outcome $k$ using operators $A_{k}$ in
the pre- and post-selected ensemble $\langle\phi||\psi\rangle$ is
given by eq. (\ref{Pkprepost}): 
\begin{equation}
P_{A}(k|\phi\psi)=\frac{|\langle\phi|A_{k}|\psi\rangle|^{2}}{\sum_{k}|\langle\phi|A_{k}|\psi\rangle|^{2}}\ .\label{eq:TheoremPA-1}
\end{equation}
Then we have:

\emph{Theorem 1: For any subnormalised rank one POVM $M_{k}$, there
exists a subnormalised CP map $A_{k}$, such that $P_{A}(k|\phi\psi)=P_{M}(k|\overline{\phi}\psi)$.
Conversely, for any subnormalised CP map $A_{k}$, there exists a
subnormalised POVM $M_{k}$, such that $P_{M}(k|\overline{\phi}\psi)=P_{A}(k|\phi\psi)$.}

This result, combined with the faxt that rank 1 POVM's are always
the most informative (see end of section \ref{sub:State-est-with-fixed-postselectedstate}),
shows that the problem of estimating the unknown pre-selected state
$|\overline{\phi}\rangle|\psi\rangle$ in the presence of a fixed
post-selected state is completely equivalent to estimating the pre-
and post-selected state $\langle\phi||\psi\rangle$ in the presence
of a fixed post-selected state.

In the case where there is no fixed post-selected state, we have implication
in one direction only:

\emph{Theorem 2: For any normalised rank one POVM $M_{k}$ ($\sum_{k}M_{k}=\one$),
there exists a normalised CP map $A_{k}$ ($\sum_{k}A_{k}^{\dagger}A_{k}=\one$),
such that $P_{A}(k|\phi\psi)=P_{M}(k|\overline{\phi}\psi)$ . }

This result shows that the problem of estimating the unknown pre-selected
state $|\overline{\phi}\rangle|\psi\rangle$ without any fixed post-selection
is always at least as hard as estimating the pre- and post-selected
state $\langle\phi||\psi\rangle$ (without any fixed post-selection). 

One would expect that the converse of Theorem 2 should not hold, since
the presence of some post-selection should give additional discriminating
power. Below we show that this intuition is correct, and provide an
example showing that the converse of Theorem 2 does not hold, i.e.
in some cases estimating the unknown pre-selected state $|\overline{\phi}\rangle|\psi\rangle$
without any fixed post-selection is harder than estimating the pre-
and post-selected state $\langle\phi||\psi\rangle$ without any fixed
post-selection.

\subsection{Proof of Theorems.}

\emph{Proof of Theorem 1.}

\emph{Part 1.} Consider the rank 1 subnormalised POVM $M_{k}=|m_{k}\rangle\langle m_{k}|$.
We will construct the Kraus operators $A_{k}$ so that the probabilities
of outcomes of measurement $A_{k}$, $P_{A}(k|\phi\psi)$, are identical
to the probabilities of outcomes of the measurement $M$: $P_{A}(k|\phi\psi)=P_{M}(k|\overline{\phi}\psi)$.

Let us rewrite 
\begin{equation}
|\langle m^{k}|\overline{\phi}\rangle|\psi\rangle|^{2}=|\sum_{\alpha\beta}m_{\alpha\beta}^{k}\langle\alpha|\overline{\phi}\rangle\langle\beta|\psi\rangle|^{2}
\end{equation}
 where $m_{\alpha\beta}^{k}$are the coefficients of $|m_{k}\rangle$
in basis $|\alpha\rangle\otimes|\beta\rangle$, 
\begin{equation}
m_{\alpha\beta}^{k}=\langle m_{k}||\alpha\rangle\otimes|\beta\rangle\ ,\label{eq:Theomk-1}
\end{equation}
 and $|\alpha\rangle$ is the basis in which complex conjugation of
$|\phi\rangle$ is defined. Let us now consider the Kraus operators
\begin{equation}
A^{k}=\sum_{\alpha\beta}|\alpha\rangle\langle\beta|A_{\alpha\beta}^{k}\label{eq:TheoAk-1}
\end{equation}
 with the choice 
\begin{equation}
A_{\alpha\beta}^{k}=m_{\alpha\beta}^{k}/\sqrt{d}\ ,\label{eq:choiceAk}
\end{equation}
where $d$ is the dimension of the Hilbert space of state $|\phi\rangle$.
(The reason for this choice of normalisation will appear below). We then have 
\begin{equation}
|\langle\phi|A_{k}|\psi\rangle|^{2}=\langle\overline{\phi}|\otimes\langle\ \psi|M_{k}|\overline{\phi}\rangle\otimes|\psi\rangle/d
\end{equation}
 Inserting this identity into eqs. (\ref{eq:TheoremPM-1},\ref{eq:TheoremPA-1})
proves the equality $P_{A}(k|\phi\psi)=P_{M}(k|\overline{\phi}\psi)$.

Note that we have $\sum_{k\alpha}\overline{A}_{\alpha\beta'}^{k}A_{\alpha\beta}^{k}=\sum_{k\alpha}\overline{m}_{\alpha\beta'}^{k}m_{\alpha\beta}^{k}/d$.
Using the subnormalisation $\sum_{k}M_{k}\leq\one$, and the fact that partial trace preserves inequalities between matrices (that is if $A$ and $B$ act on $H\otimes H'$, and $A\leq B$, then ${\rm tr}_H A \leq {\rm tr}_H B$), we have 
\begin{equation}
\sum_{k\alpha}\overline{A}_{\alpha\beta'}^{k}A_{\alpha\beta}^{k}\leq\sum_{\alpha}\delta_{\beta\beta'}/d=\delta_{\beta\beta'}\label{eq:normAk}
\end{equation}
(where the inequality is taken to be a matrix inequality, not an inequality for each $\beta \beta'$).
This implies that the Kraus operators are also subnormalized $\sum_{k}A_{k}^{\dagger}A_{k}\leq\one$.

\emph{Part 2.} Consider the subnormalised Kraus operators $A_{k}$.
We will construct a rank one POVM $M_{k}=|m_{k}\rangle\langle m_{k}|$
such that the probabilities of outcomes of measurement $M_{k}$, $P_{M}(k|\overline{\phi}\psi)$,
are identical to the probabilities of outcomes of the measurement
$A_{k}$: $P_{M}(k|\overline{\phi}\psi)=P_{A}(k|\phi\psi)$. The argument
is essentially the reverse of the argument given in Part1. We write
the Kraus operators and POVM elements using the notation of eqs. (\ref{eq:TheoAk-1},\ref{eq:Theomk-1})
and choose the $m_{\alpha\beta}^{k}$ according to 
\begin{equation}
m_{\alpha\beta}^{k}=cA_{\alpha\beta}^{k}\ .
\label{mcA}
\end{equation}
 where $c>0$ is a constant we will fix below. With this choice we
have 
\begin{equation}
\langle\overline{\phi}|\otimes\langle\ \psi|M_{k}|\overline{\phi}\rangle\otimes|\psi\rangle=c^{2}|\langle\phi|A_{k}|\psi\rangle|^{2}
\end{equation}
 Inserting this identity into eqs. (\ref{eq:TheoremPM-1},\ref{eq:TheoremPA-1})
proves the equality $P_{A}(k|\phi\psi)=P_{M}(k|\overline{\phi}\psi)$.

Note that we have $\sum_{k}\overline{m}_{\alpha'\beta'}^{k}m_{\alpha\beta}^{k}=c^{2}\sum_{k}\overline{A}_{\alpha'\beta'}^{k}A_{\alpha\beta}^{k}\geq0$
is a positive operator. By choosing $c$ sufficiently small, we can
ensure that $\sum_{k}M_{k}\leq\one$.

\emph{End of Proof.}

\emph{Proof of Theorem 2. }

Consider the rank 1 normalised POVM $M_{k}=|m_{k}\rangle\langle m_{k}|$,
$\sum_{k}M_{k}=\one$. We will construct normalized Kraus operators
$A_{k}$ so that the probabilities of outcomes of measurement $A_{k}$,
$P_{A}(k|\phi\psi)$, are identical to the probabilities of outcomes
of the measurement $M$: $P_{A}(k|\phi\psi)=P_{M}(k|\overline{\phi}\psi)$.
We proceed exactly as in the proof of Theorem 1, Part 1, and in particular
make the choice of Kraus elements eq. (\ref{eq:choiceAk}) (with the
same normalization). Then we have equality in eq. (\ref{eq:normAk})
which shows that the Kraus operators are also normalised.

\emph{End of Proof.}

\subsection{Example showing that the converse of Theorem 2 does not hold.}

The following example showing that the converse of Theorem 2 does
not hold is based on a version of the Unambiguous State Discrimination
problem.

Denote $|\psi_{\pm}\rangle=\alpha|0\rangle\pm\beta|1\rangle$ two
non orthogonal states. Denote the orthogonal states by $|\psi_{\pm}^{\perp}\rangle=\beta|0\rangle\mp\alpha|1\rangle$.
Similarly denote $|\phi_{\pm}\rangle=\sqrt{{1-\epsilon^{2}}}|0\rangle\pm\epsilon|1\rangle$
two non orthogonal states. All coefficients $\alpha>\beta>0$, $\epsilon>0$
are real, and all states are normalised: $\alpha^{2}+\beta^{2}=1$.
We will be interested in the case where $\alpha,\beta$ are fixed,
and $\epsilon$ is very small: $0\leq\epsilon\ll1$. Note that for a counter example it is in principle sufficient to consider the case when $\epsilon=0$. However this case is special since the states $|\phi_{\pm}\rangle$ are then equal and carry no information. By considering the cases when $\epsilon>0$, we show that counterexamples are rather common.

Consider the problem in which one receives either the states $|\psi_{+}\rangle|\phi_{+}\rangle$,
or the states $|\psi_{-}\rangle|\phi_{-}\rangle$. The measurement
can have three outcomes $+,-,0$: outcome $+$ can only occur if the
state was $|\psi_{+}\rangle|\phi_{+}\rangle$ (that is $P_{M}(+|-)=0$);
outcome $-$ can only occur if the state was $|\psi_{-}\rangle|\phi_{-}\rangle$
(that is $P_{M}(-|+)=0$); outcome $0$ can occur in all cases. The
aim is to minimize the probability of occurrence of outcome $0$. The
theory of USE \cite{USD1,USD2,USD3} shows without fixed post-selection,
the optimal discrimination probabilities are $P_{M}(+|+)=P_{M}(-|-)=\sqrt{1-(\alpha^{2}-\beta^{2})^{2}(1-2\epsilon^{2})^{2}}=\sqrt{1-(\alpha^{2}-\beta^{2})^{2}}+O(\epsilon^{2})$
where we expand to first order in $\epsilon^{2}$, and $P_{M}(0|+)=P_{M}(0|-)=1-\sqrt{1-(\alpha^{2}-\beta^{2})^{2}}+O(\epsilon^{2})$.

Now consider the related problem where some of the information is
flowing from the future. The aim is to distinguish between the two
ensembles $\langle\phi_{+}||\psi_{+}\rangle$ and $\langle\phi_{-}||\psi_{-}\rangle$.
The measurement, given by Kraus operators, can have three outcomes
$+,-,0$: outcome $+$ can only occur if the ensemble is $\langle\phi_{+}||\psi_{+}\rangle$
(that is $P_{A}(+|-)=0$); outcome $-$ can only occur if the ensemble
is $\langle\phi_{-}||\psi_{-}\rangle$ (that is $P_{A}(-|+)=0$);
outcome $0$ can occur in all cases. The aim is to minimize the probability
of occurrence of outcome $0$. To this end we consider the Kraus operators
$A_{+}=|0\rangle\langle\psi_{-}^{\perp}|/\sqrt{{2\alpha^{2}}}$, $A_{-}=|0\rangle\langle\psi_{+}^{\perp}|/\sqrt{{2\alpha^{2}}}$,
$A_{0}=|1\rangle\langle0|\sqrt{{1-\beta^{2}/\alpha^{2}}}$. One checks
that $A_{+}^{\dagger}A_{+}+A_{-}^{\dagger}A_{-}+A_{0}^{\dagger}A_{0}=\one$.
One easily computes that $P_{A}(+|+)=P_{A}(-|-)=1-O(\epsilon^{2})$
and $P_{A}(0|+)=P_{A}(0|-)=O(\epsilon^{2})$. 

Thus in this example, in the absence of fixed post-selection, the
outcome $0$ occurs with probability $P_{M}(0|+)=P_{M}(0|-)=O(1)$
when all the information comes from the past, and occurs with probability
$P_{A}(0|+)=P_{A}(0|-)=O(\epsilon^{2})$ when some of the information
flows from the future. The gain is dramatic. The origin of the gain
is that the states $|\phi_{\pm}\rangle$ contain very little information,
since $\epsilon$ is small, but when the state $\langle\phi_{\pm}|$
is post-selected, it can be used to strongly decrease the probability
of occurrence of the unwanted outcome $0$.

\section{Covariant measurements on spin 1/2 particles\label{sec:Covariant-msts-on-spin1/2}}

\subsection{Stating the problem.}

We illustrate the above formalism by the case of covariant measurements
on spin 1/2 particles. Suppose that the parameter to be estimated
is a direction uniformly distributed on the sphere: $\theta\equiv\Omega\in S_{2}$.
This direction is encoded in the pre- and post-selected state of spin
1/2 particles. The spins are polarized in direction $\Omega$, or
the opposite direction $-\Omega$. The task is to estimate the direction
$\Omega$. To each outcome $k$ of the measurement one thus associates
a guessed direction $\tilde{\Omega}(k)$. The quality of the estimate
is gauged with the Fidelity $F=\cos^{2}\Phi/2$ where $\Phi$ is the
angle between the true direction $\Omega$ and the guessed direction
$\tilde{\Omega}$.

When there is no post-selection the solution of this state estimation
problem is well known, see \cite{Holevo,MP,GP,M,Betal1,Betal2,Betal3}.
We summarize some of these results. Throughout this section we denote
by $N$ the total number of spins. 
\begin{enumerate}
\item \label{C1} When the initial state consists of $N$ parallel spins
$|\uparrow_{\Omega}^{\otimes N}\rangle$ the optimal fidelity is $(N+1)/(N+2)$.
\\
 
\item \label{C2} When the initial state $|\uparrow_{\Omega}^{\otimes N/2}\downarrow_{\Omega}^{\otimes N/2}\rangle$
consists of $N/2$ spins in direction $\Omega$ and $N/2$ spins in
direction $-\Omega$ (here $N$ is even), the optimal fidelity is
0.7887 for $N=2$, 0.8848 for $N=4$, 0.9235 for $N=6$. \\
 
\item \label{C3} There is an optimal encoding of the direction $\Omega$
into states of the form $R_{\Omega}|\psi\rangle$ where $R_{\Omega}$
is the rotation that maps direction $+z$ onto direction $\tilde{\Omega}$.
In the case of $N$ spins, the optimal fidelity for the optimal choice
of $\psi$ is 0.7887 for $N=2$, 0.8873 for $N=4$, 0.9306 for $N=6$. 
\end{enumerate}
The standard approach to these estimation problems is to use covariant
measurements. By covariant measurements we mean that there exists
a POVM element $M_{\tilde{\Omega}}$ for each possible guessed direction
$\tilde{\Omega}\in S_{2}$. These POVM elements are related to each
other by $M_{\tilde{\Omega}}=R_{\tilde{\Omega}}M_{\tilde{z}}R_{\tilde{\Omega}}^{\dagger}$
where $R_{\tilde{\Omega}}$ is the rotation that maps direction $+z$
onto direction $\tilde{\Omega}$ and $M_{\tilde{z}}$ is the POVM
element associated to the guessed direction $+z$.

Here we consider the problem of estimating the unknown pre-selected state $|\uparrow_\Omega^{\otimes n}\downarrow_\Omega^{\otimes m}\rangle$ in the presence of a fixed post-selected state, or the unknown pre- and post-selected ensemble
$\langle\uparrow_\Omega^{\otimes k}\downarrow_\Omega^{\otimes l}||\uparrow_\Omega^{\otimes (n-l)}\downarrow_\Omega^{\otimes(m-k)}\rangle$ in the presence of a fixed post-selected state.

Covariant measurements can also be used in the case of measurements
on pre- and post-selected ensembles. In the usual approach to state
estimation used in \cite{Holevo,MP,GP,M,Betal1,Betal2,Betal3} one
can show that covariant measurements perform at least as well as any
other measurements. We have not been able to show this in the present
case because of the more complicated form of the Fidelities. However
covariant measurements are an interesting category to consider, as
they allow for detailed calculations. Here we will restrict ourselves to covariant measurements. 
We do not know whether non-covariant
measurements could perform better for the problems considered here.

We therefore consider
subnormalized POVM elements that are related
through $M_{\tilde{\Omega}}=R_{\tilde{\Omega}}M_{\tilde{z}}R_{\tilde{\Omega}}^{\dagger}$,
or subnormalized Kraus operators that are related through $A_{\tilde{\Omega}}=R_{\tilde{\Omega}}A_{\tilde{z}}R_{\tilde{\Omega}}^{\dagger}$,
where $R_{\tilde{\Omega}}$ is the rotation that maps direction $+z$
onto direction $\tilde{\Omega}$.

\subsection{Covariant measurements and the equivalence between information flowing from the past and future.}

We note that spin 1/2 states pointing in opposite directions are related through convex conjugation and the action of a fixed unitary: $|\downarrow_\Omega\rangle= i \sigma_y |\overline{\uparrow_\Omega}\rangle$. Therefore Theorems 1 and 2 apply. We also expect Theorems 1 and 2 to apply if we restrict ourselves to covariant measurements. We now show that this is indeed the case:

\emph{Theorem 3: The relations and equivalences between estimation of pre-selected ensembles and pre- and post-selected ensembles expressed in Theorems 1 and 2 also hold if one considers covariant measurements (as defined above) 
on the ensembles $|\uparrow_\Omega^{\otimes n}\downarrow_\Omega^{\otimes m}\rangle$ and
$\langle\uparrow_\Omega^{\otimes k}\downarrow_\Omega^{\otimes l}||\uparrow_\Omega^{\otimes (n-l)}\downarrow_\Omega^{\otimes(m-k)}\rangle$, for any $k,l$, with $n,m$ fixed.}

\emph{Proof of Theorem 3. }
The proof follows easily from the proofs of Theorems 1 and 2.

Note that without changing the state estimation problem we can consider the equivalent ensembles
$|\uparrow_\Omega^{\otimes n}\overline{\uparrow_\Omega^{\otimes m}}\rangle$ and
$\langle\uparrow_\Omega^{\otimes k}\overline{\uparrow_\Omega^{\otimes l}}||\uparrow_\Omega^{\otimes (n-l)}\overline{\uparrow_\Omega^{\otimes(m-k)}}\rangle$ since they differ from the original ensemble only by fixed unitaries.

A covariant rank 1 POVM element on the above state has the form
$M_{\tilde{\Omega}}=|m_{\tilde{\Omega}}\rangle\langle m_{\tilde{\Omega}}|$
with 
\begin{equation}
|m_{\tilde{\Omega}}\rangle=\left( U_{\tilde{\Omega}} \right)^{\otimes n}
\left( {\overline{U_{\tilde{\Omega}}}} \right)^{\otimes m}|m_{\tilde{z}}\rangle
\label{mcov}\end{equation}
with $U_{\tilde{\Omega}}$ the $2\times 2$ matrix that takes a spin 1/2 pointing in the $z$ direction to the $\tilde{\Omega}$ direction.
Similarly a covariant Kraus operators acting on the above state has the form
\begin{equation}
A_{\tilde{\Omega}}=\left( U_{\tilde{\Omega}} \right)^{\otimes k}
\left(  {\overline{U_{\tilde{\Omega}}}} \right)^{\otimes l}
A_{\tilde{z}}\left( U_{\tilde{\Omega}}^\dagger \right)^{\otimes (n-l)}
\left(  {\overline{U_{\tilde{\Omega}}}^\dagger} \right)^{\otimes (m-k)}\ .
\label{Acov}\end{equation}

The key to the proofs of Theorems 1 and 2 are the mappings eqs. (\ref{eq:choiceAk}) and (\ref{mcA}) between rank 1 POVM elements and Kraus operators. It is easy to see by direct substitution that these mappings conserve the covariant character of the measurements. That is if we take a covariant rank 1 POVM element of the form eq. (\ref{mcov}) and insert it in eq. (\ref{eq:choiceAk}) we obtain a covariant Kraus operator of the form eq. (\ref{Acov}). And similarly, if we take a covariant Kraus operator of the form eq. (\ref{Acov}) and insert it in eq. (\ref{mcA}) we obtain a covariant rank 1 POVM element of the form eq. (\ref{mcov}).
\emph{End of proof. }

\subsection{Pre- selected parallel spins and fixed post-selected state.}

\label{VA}

We now discuss two examples involving pre-selected ensembles of spin 1/2 particles with fixed post-selection. 
In the first example we have obtained an analytical
result for arbitrary number $N$ of spins, while for the example of subsection \ref{VB}
we have had to resort to symbolic a math program, and have only obtained
(numerical) results for $N\leq6$ spins. We discuss the calculations
for the first example in detail, and treat the second example
more succinctly.

In this subsection we consider the case where the spins are pre-selected in
the state $|\uparrow_{\Omega}^{\otimes N}\rangle$ and there is a
fixed post-selected state $\langle0|$. We describe the different
steps of the calculation in detail. The fidelity can be expressed
as: 
\begin{equation}
F_{||}^{pre}=\frac{1}{4\pi}\int d\Omega\frac{\int d\tilde{\Omega}\ \langle\uparrow_{\Omega}^{\otimes N}|R_{\tilde{\Omega}}M_{\tilde{z}}R_{\tilde{\Omega}}^{\dagger}|\uparrow_{\Omega}^{\otimes N}\rangle\cos^{2}\Phi/2}{\int d\tilde{\Omega}\ \langle\uparrow_{\Omega}^{\otimes N}|R_{\tilde{\Omega}}M_{\tilde{z}}R_{\tilde{\Omega}}^{\dagger}|\uparrow_{\Omega}^{\otimes N}\rangle}\label{eq:Fprepar-original}
\end{equation}
 where $M_{\tilde{z}}$ is the POVM acting on the spins when the guessed
direction is $+z$, normalized according to 
\begin{equation}
\int d\tilde{\Omega}\ R_{\tilde{\Omega}}M_{\tilde{z}}R_{\tilde{\Omega}}^{\dagger}\leq\one
\end{equation}

We note that we can rewrite $|\uparrow_{\Omega}^{\otimes N}\rangle\langle\uparrow_{\Omega}^{\otimes N}|=R_{\Omega}|\uparrow_{z}^{\otimes N}\rangle\langle\uparrow_{z}^{\otimes N}|R_{\Omega}^{\dagger}$
to obtain 
\begin{equation}
F_{||}^{pre}=\frac{1}{4\pi}\int d\Omega\frac{\int d\tilde{\Omega}\ \langle\uparrow_{z}^{\otimes N}|R_{\Omega}^{\dagger}R_{\tilde{\Omega}}M_{\tilde{z}}R_{\tilde{\Omega}}^{\dagger}R_{\Omega}|\uparrow_{z}^{\otimes N}\rangle\cos^{2}\Phi/2}{\int d\tilde{\Omega}\ \langle\uparrow_{z}^{\otimes N}|R_{\Omega}^{\dagger}R_{\tilde{\Omega}}M_{\tilde{z}}R_{\tilde{\Omega}}^{\dagger}R_{\Omega}|\uparrow_{z}^{\otimes N}\rangle}
\end{equation}

Note also that the integrals over $\Omega$ and $\tilde{\Omega}$
can be replaced by integrals over the whole SU(2) group using the
uniform Haar measure (since any rotation can be decomposed into a
rotation around $z$, a rotation that brings $z$ to $\Omega$, and
a rotation around $\Omega$) to obtain

\begin{eqnarray}
F_{||}^{pre} & = & \int dU\frac{\int d\tilde{U}\ \langle\uparrow_{z}^{\otimes N}|U^{\dagger}\tilde{U}M_{\tilde{z}}\tilde{U}^{\dagger}U|\uparrow_{z}^{\otimes N}\rangle\cos^{2}\Phi/2}{\int d\tilde{U}\ \langle\uparrow_{z}^{\otimes N}|U^{\dagger}\tilde{U}M_{\tilde{z}}\tilde{U}^{\dagger}U|\uparrow_{z}^{\otimes N}\rangle}\nonumber \\
 & = & \int dU\frac{\int d\tilde{U}\ \langle\uparrow_{z}^{\otimes N}|\tilde{U}M_{\tilde{z}}\tilde{U}^{\dagger}|\uparrow_{z}^{\otimes N}\rangle\cos^{2}\Phi/2}{\int d\tilde{U}\ \langle\uparrow_{z}^{\otimes N}|\tilde{U}M_{\tilde{z}}\tilde{U}^{\dagger}|\uparrow_{z}^{\otimes N}\rangle}\nonumber \\
 & = & \frac{\int d\tilde{U}\ \langle\uparrow_{z}^{\otimes N}|\tilde{U}M_{\tilde{z}}\tilde{U}^{\dagger}|\uparrow_{z}^{\otimes N}\rangle\cos^{2}\Phi/2}{\int d\tilde{U}\ \langle\uparrow_{z}^{\otimes N}|\tilde{U}M_{\tilde{z}}\tilde{U}^{\dagger}|\uparrow_{z}^{\otimes N}\rangle}\label{eq:Fprepar-simplified}
\end{eqnarray}
where in the second line we have absorbed the rotation $U$ into the
rotation $\tilde{U}$, and where in the last line we recall that $\Phi$
is the angle between the $z$ axis and the direction onto which the
$z$ axis is rotated by rotation $\tilde{U}.$ Note how the use of
covariant measurements has enabled an important simplification: in
going from eq. (\ref{eq:Fprepar-original}) to eq. (\ref{eq:Fprepar-simplified})
we have removed one integral. Equation (\ref{eq:Fprepar-simplified})
can be reexpressed as: 
\begin{equation}
F_{||}^{pre}=\frac{{\rm Tr} CM_{\tilde{z}}}{{\rm Tr}DM_{\tilde{z}}}\label{eq:FparPre}
\end{equation}
where 
\begin{equation}
C=\int d\tilde{U}\ \tilde{U}^{\dagger}|\uparrow_{z}^{\otimes N}\rangle\langle\uparrow_{z}^{\otimes N}|\tilde{U}\ \cos^{2}\Phi/2
\label{NNNN}
\end{equation}
and

\begin{equation}
D=\int d\tilde{U}\ \tilde{U}^{\dagger}|\uparrow_{z}^{\otimes N}\rangle\langle\uparrow_{z}^{\otimes N}|\tilde{U}\ .
\label{DDDD}
\end{equation}
Now recall that without loss of generality the POVM elements can be
taken to be rank one $M_{\tilde{z}}=|m_{\tilde{z}}\rangle\langle m_{\tilde{z}}|$.
Upon varying with respect to the components of $|m_{\tilde{z}}\rangle$,
one obtains the equations 
\[
C|m_{\tilde{z}}\rangle=\lambda D|m_{\tilde{z}}\rangle
\]
with 
\[
\lambda=\frac{{\rm Tr} CM_{\tilde{z}}}{{\rm Tr}DM_{\tilde{z}}}\ .
\]
 Hence the maximum Fidelity $F_{||}^{pre}$ is given by the largest
solution $\lambda$ of $\det(C-\lambda D)=0$ (compare with eq. (\ref{eq:FparPre})).

It remains to compute the matrices $C$ and $D$. To this end we note
that the vector $|\uparrow_{z}^{\otimes N}\rangle$ has total angular
momentum $S=\sqrt{\frac{N}{2}\left(\frac{N}{2}+1\right)}$, and that
under rotation the total angular momentum does not change. We can
thus restrict our analysis to the space of total angular momentum
$S=\sqrt{\frac{N}{2}\left(\frac{N}{2}+1\right)}$ whose dimension
is $N+1$. A convenient basis of this space are the eigenvectors of
$S_{z}$ which we denote $|m\rangle$, $m=-N/2,...,N/2$. 

If $U$ is the rotation that takes direction $+z$ to direction $\theta$, $\varphi$, then 
\begin{eqnarray}
U^\dagger|\uparrow_{z}^{\otimes N}\rangle
&=&\sum_{m=-N/2}^{N/2}\cos^{N/2+m}\left(\frac{\theta}{2}\right)
\sin^{N/2-m}\left(\frac{\theta}{2}\right)
\nonumber\\
& &
e^{-i(N/2-m)\varphi} \sqrt{{N \choose N/2-m}}|m\rangle
\end{eqnarray}
and $\cos \frac{\Phi}{2}=\cos\frac{\theta}{2}$. Inserting these expressions into eqs.(\ref{NNNN}) and (\ref{DDDD}),
integrating over $\varphi$ and then $\theta$ with the uniform measure over the sphere yields that the matrices $C=
\frac{m+N/2+1}{(N+1)(N+2)}\delta_{mm'}$
and $D=\frac{1}{N+1} \delta_{mm'}$ are both diagonal
in this basis. The maximum Fidelity (the largest solution of $\det(C-\lambda D)=0$)
is therefore 
\begin{equation}
\max F_{||}^{pre}=\frac{N+1}{N+2}\ .
\end{equation}
 Thus if the direction $\Omega$ is encoded into $N$ parallel spins,
then the presence of a fixed post-selected state does not help one
in estimating the direction $\Omega$, at least if we restrict ourselves
to covariant measurements.

\subsection{Pre-selected anti parallel spins and fixed post-selected state.}

\label{VB}

Let now consider the case where the spins are pre-selected to be anti-parallel,
i.e. to be in the state $|\uparrow_{\Omega}^{\otimes N/2}\downarrow_{\Omega}^{\otimes N/2}\rangle$
(for $N$ even), and there is a fixed post-selected state $\langle0|$.
In this case the fidelity reads 
\begin{eqnarray}
 & F_{anti||}^{pre}=\frac{1}{4\pi}\int d\Omega\times\nonumber \\
 & \times\frac{\int dR\ \langle\uparrow_{\Omega}^{\otimes N/2}\downarrow_{\Omega}^{\otimes N/2}|RM_{\tilde{z}}R^{\dagger}|\uparrow_{\Omega}^{\otimes N/2}\downarrow_{\Omega}^{\otimes N/2}\rangle\cos^{2}\Phi/2}{\int dR\ \langle\uparrow_{\Omega}^{\otimes N/2}\downarrow_{\Omega}^{\otimes N/2}|RM_{\tilde{z}}R^{\dagger}|\uparrow_{\Omega}^{\otimes N/2}\downarrow_{\Omega}^{\otimes N/2}\rangle}
\end{eqnarray}
 Using exactly the same reasoning as above one can bring this to the
form 
\begin{equation}
F_{anti||}^{pre}=\frac{{\rm Tr}C'M_{\tilde{z}}}{{\rm Tr}D'M_{\tilde{z}}}
\end{equation}
where 
\[
C'=\int d\tilde{U}\ \tilde{U}^{\dagger}|\uparrow_{\Omega}^{\otimes N/2}\downarrow_{\Omega}^{\otimes N/2}\rangle\langle\uparrow_{\Omega}^{\otimes N/2}\downarrow_{\Omega}^{\otimes N/2}|\tilde{U}\ \cos^{2}\Phi/2
\]
and

\[
D'=\int d\tilde{U}\ \tilde{U}^{\dagger}|\uparrow_{\Omega}^{\otimes N/2}\downarrow_{\Omega}^{\otimes N/2}\rangle\langle\uparrow_{\Omega}^{\otimes N/2}\downarrow_{\Omega}^{\otimes N/2}|\tilde{U}
\]
The maximum Fidelity is given by the largest solution $\lambda$ of
$\det(C'-\lambda D')=0$. In this case the computation of the matrices
$C'$ and $D'$ is more complicated. Using a symbolic mathematics
program, we could compute these matrices for $N=2,4,6$, yielding
for the optimal fidelities $F_{anti||}^{pre}=$0.7887 for $N=2$,
0.8873 for $N=4$, 0.9306 for $N=6$.

Thus we see that in the case of covariant measurements on anti parallel
spins, the presence of a fixed post-selected ancilla leads to a small
improvement in the fidelity (we can go from case \ref{C2} above to
the optimal fidelities case \ref{C3} above). At present we do not
understand why sometimes there is an improvement and sometimes not.

\section{Conclusion}

In summary we have raised the question of state estimation in pre-
and post-selected ensembles and set up a general formalism for this
problem. In the examples we studied we found two main processes that
play a role: 
\begin{itemize}
\item The Measurer uses the future to dump into it the results he does not
want. No attempt at all is made to use information coming from the
future. 
\item The Measurer tries to use the information from the future and no attempt
at all is made to use the future as a dump. 
\end{itemize}
In general, a measurement procedure may combine these two ideas. 

Our first general result, Theorem 1, shows that when the future can
be used to dump unwanted results, then information coming from the
future and the complex conjugate information coming from the past
are equivalent. This was illustrated by the examples involving covariant
measurements on spin 1/2 particles discussed in section \ref{sec:Covariant-msts-on-spin1/2}.
Our second general result, Theorem 2, shows that when the future cannot
be used to dump unwanted results, then information coming from the
future is always at least as good as the complex conjugate information
coming from the past. 

Obviously this is only a first study of estimating pre- and post-selected
ensembles. Our results and examples show that sometimes the presence
of a fixed post-selection or the presence of information flowing in
from the future can dramatically improve the precision with which states
can be estimated, but that in other cases the improvement is small,
or even non existent. (For instance compare the dramatic gain in \cite{A}
with the absence of gain in the example of section \ref{VA}, for
two very related state estimation problems). Future investigations
will tell us when information coming from the future can be more informative
than the complex conjugate information coming from the past), will
tell us when using the future as a dump (i.e. having a fixed post-selected
state $\langle0|$) helps and when it does not, etc...

Finally let us comment on the conceptual implications of pre- and
post-selection. The dynamics of physical systems are invariant under
time reversal. But the ``measurement postulate'' of
quantum mechanics breaks this invariance. The theory of pre- and post-selection
is an attempt to correct this and to have a theory of micro physics
that is genuinely invariant under time reversal. But as \cite{A}
and the present work show, this approach has dramatic consequences.
The hierarchy of computational complexity and much of the structure
of quantum information break down. For instance, since two states
which are arbitrarily close together can be distinguished with certainty,
an analogue of Holevo's theorem will not hold. Defining a unit of
quantum information in the pre- and post- selected setting (analog to
the usual qubit) is thus bound to be far more complicated and involve
significant conceptual steps.

We do not know what the solution to this conundrum is. Is it possible
to formulate a genuinely time invariant and satisfactory theory of
micro physics? If so how deep a reformulation of physics will it require?

\textbf{Acknowledgments.} We acknowledge financial support by EU projects
Qubit Applications (QAP contract 015848) and Quantum Computer Science
(QCS contract 255961), and by the InterUniversity Attraction Pole
-Belgium Science Policy- project P6/10 Photonics@be.


\begin{thebibliography}{References}
\bibitem{ABL} Y. Aharonov, P. G. Bergmann and J. Lebowitz, Phys.
Rev 134, B1410 (1964).

\bibitem{AV} Y. Aharonov and L. Vaidman, 
Lect. Notes Phys. 734, 395-443 (2007).

\bibitem{APTV} Y. Aharonov, S. Popescu, J. Tollaksen, and L. Vaidman,
Phys. Rev. A \textbf{79}, 052110 (2009)

\bibitem{Helstrom} C.W. Helstrom, {\em Quantum Detection and Estimation
Theory}, Academic, New York, 1976.

\bibitem{Holevo} A. S. Holevo, {\em Probabilistic and Statistical
Aspects of Quantum Theory}, North-Holland, Amsterdam, 1982.

\bibitem{MP} S. Massar and S. Popescu, Phys. Rev. Lett. 74, 1259
(1995).

\bibitem{GP} N. Gisin and S. Popescu, Phys. Rev. Lett. 83, 432 (1999).

\bibitem{M} S. Massar, Phys. Rev. A 62, 040101(R) (2000)

\bibitem{Betal1} E. Bagan, M. Baig, A. Brey, R. Munoz-Tapia and R.
Tarrach, Phys. Rev. A \textbf{63}, 052309 (2001). 

\bibitem{Betal2} E. Bagan, M. Baig and R. Munoz-Tapia, Phys. Rev.
A \textbf{64}, 022305 (2001) 

\bibitem{Betal3} E. Bagan, M. Baig, R. Munoz-Tapia, ''Communicating
a direction using spin states'', arXiv:quant-ph/0106155

\bibitem{A} S. Aaronson, 
Proc. R. Soc. A \textbf{461} (2005) 3473; arXiv:quant-ph/0412187v1

\bibitem{ref1} Y. Aharonov, D. Z. Albert, and L. Vaidman, Phys. Rev.
Lett. \textbf{60}, 1351 (1988).

\bibitem{ref4} J. S. Lundeen and A. M. Steinberg, Phys. Rev. Lett.
102, 020404 (2009); K. Yokota, T. Yamamoto, M. Koashi, and N. Imoto,
New J. Phys. \textbf{11}, 033011 (2009).

\bibitem{ref5} L. Hardy, Phys. Rev. Lett. \textbf{68}, 2981 (1992).

\bibitem{refLSPSB} J. S. Lundeen, B. Sutherland, A. Patel, C. Stewart
and C. Bamber, Nature 474, 188\textendash{}191 (2011)

\bibitem{refKBRSMSS} S. Kocsis, B. Braverman, S. Ravets, M. J. Stevens,
R. P. Mirin, L. K. Shalm and A. M. Steinberg, Science 332, 1170-1173
(2011)

\bibitem{ref6} Y. Aharonov, A. Botero, S. Popescu, B. Reznik, and
J. Tollaksen, Phys. Lett. A \textbf{301}, 130 (2002).

\bibitem{ref7} D. R. Solli, C. F. McCormick, R.Y. Chiao, S. Popescu,
and J. M. Hickmann, Phys. Rev. Lett. \textbf{92}, 043601 (2004).

\bibitem{ref8} N. Brunner, V. Scarani, M. Wegmuller, M. Legre,
and N. Gisin, Phys. Rev. Lett. \textbf{93}, 203902 (2004).

\bibitem{ref9} N. Brunner, A. Acin, D. Collins, N. Gisin, and V.
Scarani, Phys. Rev. Lett. \textbf{91}, 180402 (2003).

\bibitem{ref10} H. M. Wiseman, Phys. Rev. A \textbf{65}, 032111 (2002).

\bibitem{ref11} N.W. M. Ritchie, J. G. Story, and R. G. Hulet, Phys.
Rev. Lett. \textbf{66}, 1107 (1991); A.D. Parks, D.W. Cullin, and
D. C. Stoudt, Proc. R. Soc. A \textbf{454}, 2997 (1998); K. J. Resch,
J. S. Lundeen, and A. M. Steinberg, Phys. Lett. A \textbf{324}, 125
(2004); G. J. Pryde, J. L. O'Brien, A. G. White, T. C. Ralph, and
H. M. Wiseman, Phys. Rev. Lett. \textbf{94}, 220405 (2005); R. Mir
et al., New J. Phys. \textbf{9}, 287 (2007); M. Goggin et al., Proc.
Natl. Acad. Sci. U. S. A. \textbf{108}, 1256 (2011).

\bibitem{ref13} Y. Aharonov and L. Vaidman, Phys. Rev. A \textbf{41},
11 (1990).

\bibitem{ref14} O. Hosten and P. Kwiat, Science \textbf{319}, 787
(2008).

\bibitem{refBS} N. Brunner and C. Simon, Phys. Rev. Lett. \textbf{105},
010405 (2010)

\bibitem{refZRG} O. Zilberberg, A. Romito, and Y. Gefen, Phys. Rev.
Lett. \textbf{10}6, 080405 (2011)

\bibitem{ref15} P. B. Dixon, D. J. Starling, A. N. Jordan, and J.
C. Howell, Phys. Rev. Lett. \textbf{102}, 173601 (2009); D. J. Starling,
P. B. Dixon, A. N. Jordan, and J. C. Howell, Phys. Rev. A \textbf{80},
041803 (2009).

\bibitem{USD1} D. Dieks, Phys. Lett. A \textbf{126} 303 (1988)

\bibitem{USD2} I. D. Ivanovic, Phys. Lett. A \textbf{123} 257 (1987)

\bibitem{USD3}A. Peres, Phys. Lett. A \textbf{128} 19 (1988).

\end{thebibliography}
\end{document}